\documentclass[%
 reprint, onecolumn,
nofootinbib,
 amsmath,amssymb,
 aps,
]{revtex4-2}

\usepackage{graphicx}
\usepackage{dcolumn}
\usepackage{bm}
\usepackage{xcolor}

\usepackage[letterpaper,top=2cm,bottom=2cm,left=3cm,right=3cm,marginparwidth=1.75cm]{geometry}

\begin{document}
\preprint{APS/123-QED}

\title{Unified Framework of Forced Magnetic Reconnection and Alfv\'en Resonance}
\author{D. Urbanski}
\author{A. Tenerani}
\author{F. L. Waelbroeck}
\affiliation{
 Department of Physics, The University of Texas at Austin
}
\affiliation{Institute for Fusion Studies}
\date{\today}

\begin{abstract}
A unified linear theory that includes forced reconnection as a particular case of Alfv\'en resonance is presented. We consider a generalized Taylor problem in which a sheared magnetic field is subject to a time-dependent boundary perturbation oscillating at frequency $\omega_0$. By analyzing the asymptotic time response of the system,  the theory demonstrates that the Alfv\'en resonance is due to the residues at the resonant poles, in the complex frequency plane, introduced by the boundary perturbation. Alfv\'en resonance transitions towards forced reconnection, described by the constant-psi regime for (normalized) times $t\gg S^{1/3}$, when the forcing frequency of the boundary perturbation is $\omega_0\ll S^{-1/3}$, allowing the coupling of the Alfv\'en resonances across the neutral line with the reconnecting mode, as originally suggested in~\cite{uberoi1999alfven}. Additionally, it is shown that even if forced reconnection develops for finite, albeit small, frequencies, the reconnection rate and reconnected flux are strongly reduced for frequencies $\omega_0\gg S^{-3/5}$.  
\end{abstract}
\maketitle

\section{Introduction}

\indent

The dynamical formation of  boundary layers has long been a problem of fundamental interest to understand plasma heating in weakly-collisional magnetized plasmas such as those found in space and in most laboratory devices.  Alfv\'en resonance and forced magnetic reconnection are two energy conversion processes mediated by the formation of boundary layers and, as such, have received much attention in the past \cite{hahm1985forced,cole2004forced,wang1992forced,fitzpatrick2003numerical,comisso2015extended,tataronis1973decay,kappraff1977resistive,mok1985resistive,bertin1986alfven,hegna1999dynamics}. 

The problem of forced magnetic reconnection, known as the ``Taylor problem", was first investigated analytically by  Hahm and Kulsrud \cite{hahm1985forced} in slab geometry,  for an initial sheared magnetic field subject to a boundary perturbation. They demonstrated that the system evolves from an initial configuration that preserves the topology of the magnetic field, towards a final equilibrium with magnetic islands. This process can be described by four linear stages, with nonlinear island dynamics thereafter. The first two stages essentially correspond to the ideal evolution leading to the growth of a surface current at the neutral line that increases linearly with time. If resistivity is retained, the magnetic flux starts to reconnect at the neutral line, increasing algebraically with time. The system transitions to the resistive regime proper in the third stage, at around $t\sim S^{1/3}$, where $S$ is the Lundquist number based on the magnetic field shear length and $t$ is the time normalized to the corresponding Alfv\'en time. The long-time behavior represents the fourth stage and is described by the constant-psi regime. The constant-psi regime occurs over a time scale $t\sim S^{3/5}$, during which the reconnected flux leads to the formation of magnetic islands due to the resistive dissipation at the developed boundary layer, by ultimately reaching a stationary  equilibrium. 

Alfv\'en resonance has also been studied for a long time for its role in plasma heating via resonant absorption, with applications to the solar corona and tokamak plasmas \cite{chen1974plasma,davila1987heating,malara1994wave}. Alfv\'en resonance occurs when the wave is polarized in the plane containing the mean sheared magnetic field and its gradient.  In the presence of an inhomogeneous magnetic field, normal-mode analysis predicts a continuum spectrum of singular modes exhibiting a logarithmic singularity at the location where $\omega=k_\parallel v_a(x)$, $v_a(x)$ being the Alfv\'en speed and $k_\parallel$ the field-aligned wave vector \cite{tataronis1973decay,hasegawa1982alfven}. When resistivity is included,  such a singularity also implies that if the plasma is driven externally, a steady state is achieved where the wave energy is accumulated and absorbed at the resonant layer \cite{kappraff1977resistive}. 

In 1999, Uberoi and Zweibel \cite{uberoi1999alfven} pointed out  similarities between forced reconnection and Alfv\'en resonance, drawing from earlier studies of Alfv\'en resonance at the magnetic field neutral line \cite{uberoi1994resonant}. Indeed, the current density grows initially linearly with time for Alfv\'en resonance and forced reconnection, and resistive effects become important over a time scale $t\sim S^{1/3}$ in both processes  \cite{kappraff1977resistive, hahm1985forced}. Furthermore, they show that  both processes are described by the same  governing equation in the limit of zero frequency. Since the Alfv\'en resonance develops an inner layer $\delta$ that scales as $\delta\sim S^{-1/3}$ \cite{mok1985resistive}, it was reasonably argued that the transition to forced reconnection should occur for values of the frequency of the injected wave $\omega\sim S^{-1/3}$. However, an explicit time-dependent solution to the boundary value problem, demonstrating that Alfv\'en resonance and forced magnetic reconnection are in fact intrinsically related, has not yet been determined. 

The purpose of this paper is to provide a unified theory that includes forced reconnection as a particular case of Alfv\'en resonance. We solve the time dependent, boundary value problem for both the early and long time stages of the evolution of the system, by assuming a perturbation of arbitrary frequency $\omega_0$ away from the neutral line along the lines of the Taylor problem.

The paper is organized as follows: in section 2 we present the model equations; in section 3 and 4 we derive the most general solution to the time-dependent boundary value problem (in Laplace and Fourier space) for arbitrary boundary conditions. The solution is derived by matching the so-called outer solution (section 3) to the inner layer solution (section 4) by applying the asymptotic matching technique; in section 5 and 6 we consider the special case of a standing wave of arbitrary frequency as boundary condition. We derive, analytically or by inverting our solutions to time and configuration space with Mathematica~\cite{Mathematica}, the explicit space-time solution in the early-time stage  (section 5), and in the time-asymptotic stage (section 6). In section 6,  we discuss how the known solutions for forced reconnection and Alfv\'en resonance are both recovered by analyzing in the complex plane the dominant contributions to the stream function, and we compare our solutions with  linear 2D simulations described in the appendix. Finally, we discuss our results and provide a summary in section~7.   

\section{Model equations and boundary conditions}

We consider the two-dimensional resistive magnetohydrodynamic (MHD) equations for an inviscid and incompressible plasma in slab geometry. In this model, the unperturbed state is represented by a stationary  plasma with uniform density $\rho$ and anti-symmetric sheared  magnetic field given by 
\begin{eqnarray}
    \boldsymbol{B_0}(x) = {B}_{0} \frac{x}{a} {\bf \hat{y}},
    \label{eq:Background Mag}
\end{eqnarray}
as is customary for the Taylor problem \cite{hahm1985forced,cole2004forced,comisso2015extended}. An inhomogeneous pressure or out-of-plane magnetic field are understood to maintain the equilibrium. The system is then perturbed at $t=0$ by a time-dependent displacement of the boundary at $x=\pm a$, represented by a forcing function $\Xi(y,t)$. The resulting  perturbed magnetic and velocity fields are defined in terms of the flux ($\psi$) and stream ($\phi$) functions given, respectively, by
\begin{equation}
\boldsymbol{B} = \nabla \psi \times {\bf \hat{z}},
\end{equation}
\begin{equation}
\boldsymbol{v} = \nabla \phi \times {\bf \hat{z}}.
\end{equation}

To solve the time dependent, initial value problem for $\psi$ and $\phi$, we perform both a Fourier transform with respect to the  $y$ variable, and a Laplace transform in time. The stream and flux functions, $\phi$ and $\psi$, are zero at $t = 0$ and the driver is provided by the boundary condition. We obtain the following set of linearized equations,
\begin{equation}
    g(\hat{\phi}'' -k^2\hat{\phi}) = ikx(\hat{\psi}'' -k^2\hat{\psi}),
    \label{eq:Scalar Eq of Motion}
\end{equation}
\begin{equation}
    g\hat{\psi} = \frac{1}{S}(\hat{\psi}''-k^2\hat{\psi}) + ikx\hat{\phi},
    \label{eq:Scalar Induction}
\end{equation}
where prime denotes the derivative with respect to $x$, and $g$ and $k$ are the Laplace and Fourier variables defined  with the following conventions,
\begin{equation}
    \hat{f}(g) = \mathcal{L}(f) = \int_0^\infty f(t)e^{-g t}dt,
    \label{eq:Laplace Transform}
\end{equation}

\begin{equation}
    \bar{f}(k) = \mathcal{F}(f) = \frac{1}{2\pi}\int_{-\infty}^{\infty}f(y)e^{-iky}dy,
    \label{eq:Fourier Conventions}
\end{equation}
for an arbitrary function $f$. For ease of notation, hereafter we will denote with hats Laplace transforms (therefore functions of $g$), which may or may not depend on the Fourier variable $k$. In case of ambiguity, we will indicate explicitly the variables upon which a given function depends. Here, length scales have been normalized to the slab width $a$, and time to the Alfv\'en time defined by $\tau_a = a/v_a$, where $v_a=B_0/\sqrt{4\pi\rho}$ is the Alfv\'en speed. The Lundquist number $S=av_a/\eta$ is the ratio of the resistive times scale $\tau_r =  a^2/\eta$ to $\tau_a$, $\eta$ being the magnetic diffusivity. 

We look for solutions to Eqs.~(\ref{eq:Scalar Eq of Motion})-(\ref{eq:Scalar Induction}) in the domain $x=[-1,1]$, subject to a displacement of the boundaries  $\hat\Xi_\pm(k, g)$. The time-dependent solution for a given Fourier mode will then be obtained by taking the inverse Laplace transform. The form of the boundary displacement (the forcing) will be left general for much of the derivation, but it is assumed periodic in the azimuthal $y$ direction and symmetric about the midplane $x=0$. The general boundary conditions that complete our problem  are therefore represented~by
\begin{equation}
    \hat{\psi}(\pm 1, k, g) = \hat{\Xi}(k,g),
    \label{eq:Scalar psi BC}
\end{equation}
\begin{equation}
    \hat{\phi}(\pm 1, k, g) = \pm \frac{g}{i k}\hat{\Xi}(k,g),
    \label{eq:Scalar phi BC}
\end{equation}
with Eq.~(\ref{eq:Scalar phi BC})  following from the ideal limit of Eq.~\eqref{eq:Scalar Induction}.

\section{The Time Dependent Boundary Layer Problem}
Essential to the theories of both forced magnetic reconnection and Alfv\'en resonance is the formation of boundary layers about their resonant surfaces. We assume that boundary layers develop for $|x|\ll 1$. This means that the regime we are exploring for the Alfv\'en resonance is that of a sub-Alfv\'{e}nic resonant frequency  $\omega_0\tau_a\ll 1$. Therefore, for $S\gg1$, the plasma can be assumed to be governed by ideal MHD (or $S=\infty$) except in the immediate vicinity of the boundary layer, which contains large gradients. In the boundary layer, or inner layer, resistive MHD must be considered, however, because of the large gradients, one can assume that $\partial^2/\partial x^2\gg  k^2$. In this way, equations \eqref{eq:Scalar Eq of Motion} and \eqref{eq:Scalar Induction} can be solved by breaking the domain into two regions: an ideal ``outer region" away from the boundary layer, and an ``inner layer" which contains all the physics of the problem but is treated as one dimensional. These regions are then asymptotically matched for a fully general solution valid in the entire domain.

\subsection{The Outer Region}

The region away from the boundary layer where gradients are small composes much of the plasma and is treated as both ideal and in  steady state. Equations \eqref{eq:Scalar Eq of Motion} and \eqref{eq:Scalar Induction} become, under these conditions, 
\begin{eqnarray}
    \hat{\psi}''= k^2\hat{\psi},
    \label{eq:Outer Eq of Motion}
\\
    \hat{\phi} = \frac{g}{i k x}\hat{\psi}.
    \label{eq:Outer Induction}
\end{eqnarray}
The general solution to Eq.~\eqref{eq:Outer Eq of Motion} that  satisfies the boundary condition given by  Eq.~\eqref{eq:Scalar psi BC} and is even in $x$ is given by \cite{hahm1985forced}
\begin{equation}
    \hat{\psi}^{out} = \hat{\psi}_0( k, g) \left[\cosh(k x) - \frac{\sinh(k|x|)}{\tanh(k)}\right] + \hat{\Xi}(k,g) \frac{\sinh(k|x|)}{\sinh(k)},
    \label{eq:Psi Outer}
\end{equation}
where $\hat\psi_0(k,g)=\hat\psi^{out}(x=0)$. The outer solution for $\hat\phi$ follows directly from Eq.~\eqref{eq:Outer Induction}. The solution (\ref{eq:Psi Outer}) consists of two contributions: a contribution satisfying homogeneous boundary conditions (no forcing) that takes a time dependent value of amplitude $\hat{\psi}_0( k, g)$ at the rational surface $x=0$ and an ideal contribution describing a ``screened'' plasma response that vanishes at $x=0$ that satisfies the boundary condition. As can be seen from Eq. \eqref{eq:Psi Outer},  $\hat{\psi}^{out}$ exhibits a jump in its derivative which we define here as
\begin{eqnarray}
    \Delta\Psi' \equiv -\frac{2 k}{\tanh(k)}\hat{\psi}_0 + \frac{2k}{\sinh(k)}\hat{\Xi}.
    \label{eq:Delta Psi}
\end{eqnarray}
The first term on the right-hand-side of Eq.~\eqref{eq:Delta Psi} contains the  $\Delta^\prime$ parameter that determines the stability to tearing mode of an unforced current sheet equilibrium \cite{furth1963finite}, which is negative in this case (thus, the equilibrium defined by Eq.~\eqref{eq:Background Mag} is stable to tearing mode). The second term represents the effect of the forcing through the function $\hat\Xi$.

Close to the origin, the outer solutions for the flux and stream functions, under series expansion, tend to %
\begin{equation}
    \hat{\psi}^{out}(x\rightarrow0) = \hat{\psi}_0 + \frac{1}{2}\Delta\Psi'|x|,
    \label{eq:Psi Outer Limit 0}
\end{equation}
\begin{eqnarray}
    \hat{\phi}^{out}(x\rightarrow0) = \frac{g}{ikx}\hat{\psi}_0\left(1+ \frac{1}{2}\frac{\Delta\Psi'}{\hat{\psi}_0} |x|\right).
    \label{eq:Phi Outer Limit 0}
\end{eqnarray}
The limit of the outer solution for $ x\rightarrow 0$ expressed by Eq.~\eqref{eq:Psi Outer Limit 0} and \eqref{eq:Phi Outer Limit 0}, provides the boundary condition for the inner layer solution as it tends away from the resonance. We anticipate that the inner layer equations can be solved  for $\hat\phi$ by applying a Fourier transform with respect to the variable $x$  \cite{pegoraro1989internal,porcelli1987viscous}. With that in mind, it follows from Eq.~\eqref{eq:Phi Outer Limit 0} that the asymptotic behavior of the Fourier transform of ${\phi}^{out}(x)$, denoted as 
$\Tilde{\phi}^{out}(\theta)$,  at large 
$\theta$ is given by 
\begin{eqnarray}
    \Tilde{\phi}^{out}(\theta\rightarrow\infty) = -\frac{g}{k}\frac{\hat{\psi_0}}{2}\frac{1}{\theta}\left(|\theta|+\frac{1}{\pi}\Delta_g\right),
    \label{eq:Phi Outer Limit Theta}
\end{eqnarray}
where from now on we use the tilde $\Tilde{[\ ]}$ to denote functions in $\theta$ space, and 
\begin{equation}
    \Delta_g \equiv \frac{\Delta\Psi'}{\hat{\psi}_0}
\end{equation}
is the layer response. The asymptotic form in Eq.~(\ref{eq:Phi Outer Limit Theta}) must be matched with the inner layer solution.

\subsection{The Inner Layer}

In the inner layer, where gradients are large and resistive effects must be included, equations \eqref{eq:Scalar Eq of Motion} and \eqref{eq:Scalar Induction} are approximated to 
\begin{eqnarray}
    g\hat{\phi}'' = ikx\hat{\psi}'',
    \label{eq:Inner Eq of Motion}
\\
    \hat{\psi} = \frac{1}{ g S}\hat{\psi}'' + \frac{ikx}{g}\hat{\phi},
    \label{eq:Inner Induction}
\end{eqnarray}
respectively. Just like equations \eqref{eq:Scalar Eq of Motion}-\eqref{eq:Scalar Induction}, the system of  equations \eqref{eq:Inner Eq of Motion}-\eqref{eq:Inner Induction} admit solutions for $\hat\psi$ and $\hat \phi$ with a defined parity. In the following we seek solutions with $\hat\psi$ even in $x$ and $\hat\phi$ odd, and we will show only the $x \geq 0$ part of the solutions. The solution in the entire domain can then be obtained by extension, according to the appropriate parity. 

To make the above set of equations analytically tractable, we Fourier transform in the $x$ direction. This approach allows one to reduce the original fourth order system of differential equations with respect to the variable $x$, to a second order system in $\theta$ that is analytically solvable \cite{pegoraro1986theory}. In this way, the differential equations in terms of $\theta$ are
\begin{eqnarray}
    \theta^2\Tilde{\phi} = -\frac{k}{g}\frac{\partial}{\partial\theta}(\theta^2\Tilde{\psi}),
    \label{eq:Inner Eq of Motion Theta}
\\
    \left(1+\frac{\theta^2}{ g S}\right)\Tilde{\psi} = - \frac{k}{g}\frac{\partial}{\partial\theta}\Tilde{\phi}.
    \label{eq:Inner Induction Theta}
\end{eqnarray}

From the definition $J_z = - \nabla^2\psi\approx \theta^2\tilde\psi$ and equation \eqref{eq:Inner Induction Theta}, we derive the following expression for $\tilde{J}_z$ in terms of~$\tilde{\phi}$,
\begin{equation}
    \Tilde{J}_z = -\frac{(Sk)^{1/3}}{Q}\frac{\theta^2}{(1+\frac{\theta^2}{(Sk)^{2/3}Q})}\frac{d}{d\theta}\Tilde{\phi},
    \label{eq:Current Density From Phi}
\end{equation}
where we have introduced the normalized Laplace variable
\begin{eqnarray}
    Q = \frac{gS^{1/3}}{k^{2/3}}.
    \label{eq:Q}
\end{eqnarray}
Next, by dividing equation \eqref{eq:Inner Eq of Motion Theta} by $\theta^2$, differentiating it with respect to $\theta$, and making use of Eq.~\eqref{eq:Current Density From Phi}, we find an expression entirely in terms of the current density 
\begin{eqnarray}
    \theta^2\frac{d}{d\theta}\left(\frac{(Sk)^{2/3}}{\theta^2}\frac{d}{d\theta}\Tilde{J_z}\right) =  Q^2 \left(1+\frac{\theta^2}{(Sk)^{2/3}Q}\right)\Tilde{J_z}.
    \label{eq:Current Density Eq in Theta}
\end{eqnarray}
Finally, with the change of variable 
\begin{equation}
    z = \frac{\sqrt{Q}\theta^2}{(Sk)^{2/3}},
    \label{eq:z}
\end{equation}
and with the  ansatz $\Tilde{J}_z(z)= e^{-\frac{z}{2}} \tilde{f}(z)$ (based on the expected behavior of the solution by taking the limit $z\gg1$ of Eq.~\eqref{eq:Current Density Eq in Theta}), Eq. \eqref{eq:Current Density Eq in Theta} takes the form of Kummer's confluent hypergeometric equation for the function $\tilde{f}(z)$:
\begin{equation}
    z \tilde{f}'' + \left(-\frac{1}{2} - z\right)\tilde{f}' - \frac{(Q^{3/2} - 1)}{4}\tilde{f} = 0.
\end{equation}

The solution to Kummer's equation that is well behaved for $z\rightarrow\infty$ \cite{pegoraro1986theory} is the Tricomi confluent hypergeometric function $U(a,b,z)$ \cite{abramowitz1988handbook}. In terms of this function, the solution to Eq.~\eqref{eq:Current Density Eq in Theta} for the current density takes the form 
\begin{equation}
    \Tilde{J}_z = A e^{-\frac{z}{2}}U\left(\frac{(Q^{3/2}-1)}{4},-\frac{1}{2},z\right),
    \label{eq:Jz General z}
\end{equation}
where $A$ is an amplitude to be determined via asymptotic matching. By using the recurrence properties of the exponential and of the function $U(a,b,z)$, together with equation \eqref{eq:Inner Eq of Motion Theta} upon substitution of $\tilde{J}_z = \theta^2\tilde{\psi}$, we determine the inner layer solution for $\Tilde{\phi}$ as:
\begin{equation}
    \Tilde{\phi}^{in} = \frac{2 A}{(Sk)^{2/3}Q^{1/4}}\frac{e^{-\frac{z}{2}}}{\sqrt{z}}
     \left[U\left(\frac{(Q^{3/2}-1)}{4},\frac{1}{2},z\right)-\frac{1}{2}U\left(\frac{(Q^{3/2}-1)}{4},-\frac{1}{2},z\right) \right].
    \label{eq:Phi Inner}
\end{equation}

In the limit of small $\theta$ (or large $x$), the solution to be asymptotically matched with Eq. \eqref{eq:Phi Outer Limit Theta} is
\begin{equation}    
\Tilde{\phi}^{in}_{\theta\rightarrow0} = \frac{2\sqrt{\pi}A}{(Sk)^{2/3}Q^{1/4}}\frac{1}{\theta}\left(\frac{Q^{5/4}}{(Q^{3/2}+1)}\frac{(Sk)^{1/3}}{\Gamma\left(\frac{Q^{3/2}+1}{4}\right)} - \frac{2}{\Gamma\left(\frac{Q^{3/2}-1}{4}\right)}\theta\right)
    \label{eq:Phi Inner Asymptotic}
\end{equation}
with $\Gamma$ the Euler Gamma function.

\section{The Matched Inner Layer Solution}

The most general matched inner layer solution is determined by Eq. $\eqref{eq:Phi Inner}$ with the following amplitude $A$ determined by asymptotic matching,
\begin{equation}
    A = \frac{\hat{\psi}_0(k,g)}{8\sqrt{\pi}}Q^{5/4}(Sk)^{1/3}\Gamma\left( \frac{(Q^{3/2}-1)}{4}\right),
    \label{eq:A}
\end{equation}
with
\begin{equation}\label{eq:Psi0 Def}
    \hat{\psi}_0(k,g) = \hat{\Xi}(k,g)\frac{\frac{2k}{\sinh[k]}}{\Delta_g+\frac{2k}{\tanh[k]}} \equiv \hat{\Xi}(k,g)\frac{\Delta_f}{\Delta_g - \Delta_0},
\end{equation}
and
\begin{equation}\label{eq:Delg}
     \Delta_g = -2\pi (Sk)^{1/3} \frac{Q^{5/4}}{(Q^3-1)}\frac{\Gamma\left(\frac{(Q^{3/2}-1)}{4}+1\right)}{\Gamma\left(\frac{(Q^{3/2}-1)}{4}+\frac{1}{2}\right)}.
\end{equation}
Here $\Delta_f = \frac{2k}{\sinh{k}}$ and $\Delta_0 = -\frac{2k}{\tanh{k}}$, with $\Delta_0$ being the conventional jump of the derivative of $\hat\psi^{out}$ at the origin (the $\Delta^\prime$ parameter of the tearing mode).

Equation \eqref{eq:Phi Inner}, together with Eqs. \eqref{eq:A}-\eqref{eq:Delg}, completely describes the stream function for both forced magnetic reconnection and Alfv\'en resonance, and it includes the full time evolution from the early stages, $t \ll S^{1/3}$, when the evolution is essentially ideal and the boundary layer is being formed, all the way to the long-time behavior determined by resistivity.  

We note that Eq.~\eqref{eq:Phi Inner} includes as special cases the eigenfunctions that describe the ideal kink, the resistive kink (also referred to as non constant-psi regime for reconnection in slab geometry) and the ordinary tearing mode (constant-psi  regimes) under appropriate asymptotic expansions \cite{RDR,pegoraro1986theory,CoppiGal,porcelli1987viscous}\footnote{We note a typo on equation (31) of \cite{porcelli1987viscous}.}. The important difference is that here we are solving for the time-dependent problem rather than the eigenvalue problem and, therefore, $g$ is a variable; additionally, we include the effect of a general forcing function at the boundary through the function $\hat{\Xi}$ by allowing not only (forced) reconnection, but also Alfv\'en resonances.

The solution given in Eq.~\eqref{eq:Phi Inner} will now be considered for two distinct stages of the dynamical evolution of the system. The early-time ideal regime corresponds to the limit $Q\gg1$ ($t\ll S^{1/3}$). The time-asymptotic resistive regime, when resistivity has allowed for the formation of a finite width boundary layer, corresponds to the limit $Q\ll1$ ($ t\gg S^{1/3}$). These two stages, which forced reconnection and Alfv\'en resonance have in common, are naturally recovered through two asymptotic expansions of the general solution \eqref{eq:Phi Inner} with respect to the parameter $Q$. The transition between the ideal and the resistive regimes occurs for $Q\sim 1$, or for times $t\sim S^{1/3}$. Thus, with Eq.~\eqref{eq:Phi Inner} we recover the expected time scales.

To make progress, and to provide a specific solution to discuss how forced reconnection and Alfv\'en resonances are recovered from Eq.~\eqref{eq:Phi Inner} in the two aforementioned stages, we will consider a specific boundary forcing given by a standing wave of the following form, 
\begin{equation}
    \Xi (y, t)= \Xi_0\cos(\omega_0 t)\cos(k_0 y),
    \label{eq:forcing}
\end{equation}
of small amplitude $\Xi_0$. The Fourier and Laplace transformed boundary condition is, therefore,
\begin{equation}
    \hat{\Xi}( k, g) = \frac{\Xi_0}{2} \frac{g}{(g^2+\omega_0^2)} \left[\delta(k-k_0) +  \delta(k+k_0)\right].
    \label{eq:forcing_g}
\end{equation}
Even though we specialize our results to the forcing in \eqref{eq:forcing}, the validity of our discussion holds in general.

\section{The Ideal Regime}
\begin{figure}
    \centering
    \includegraphics[width=.4\textwidth]{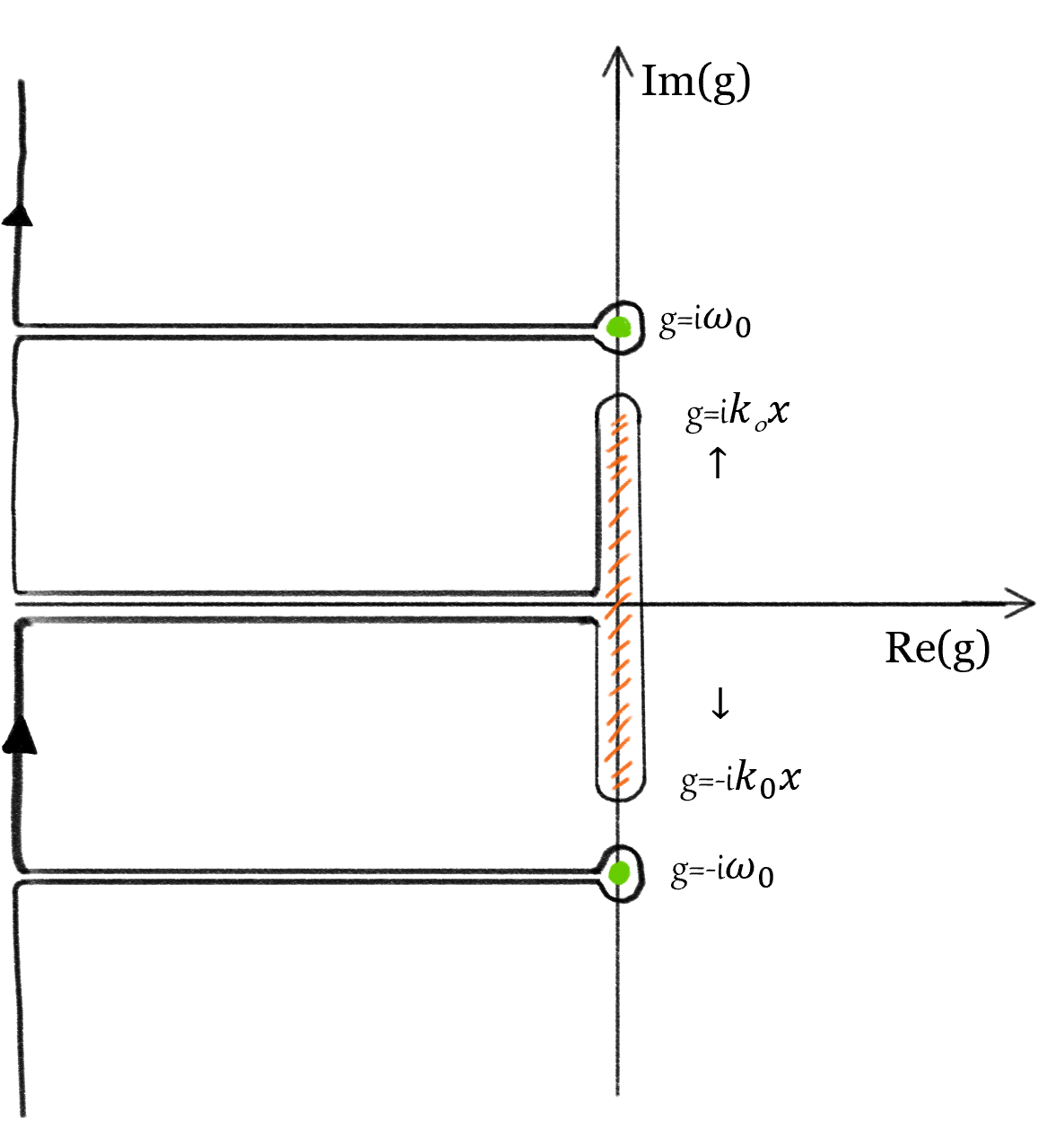}
    \caption{The branches and poles of the ideal solution \eqref{eq:Phi Ideal x} for the standing wave boundary forcing \eqref{eq:forcing_g} are shown here in the complex plane. In black is the Bromwich contour for the  inverse Laplace transform. The poles introduced by a finite frequency forcing determine the resonant response and are represented in green. The branch cut of the complex $\arctan$ is represented in orange and it contributes to the ideal stage of the forced reconnection solution.}
    \label{fig:Phi Ideal Contour}
\end{figure}

 The ideal solution can  be recovered from the general solution given by Eq.~\eqref{eq:Phi Inner} by taking the limit $Q\gg 1$ (or $g \gg S^{-1/3}$) and, simultaneously, $z/Q^{3/2}\ll1$. The leading order contributions, after imposing the appropriate parity for the stream function,  yield

\begin{eqnarray}
    \Tilde{\phi}(\theta,k, g) = \frac{\hat{\psi}_0}{2}\frac{e^{ -\frac{g}{|k|} \theta}}{\theta}.
    \label{Phi Ideal Theta}
\end{eqnarray}
Under this limit, $\Delta_0\ll\Delta_g$ and is thus neglected. Then, $\hat{\psi}_0$ takes the form

\[ 
    \hat{\psi}_0 = \hat{\Xi}\frac{\Delta_f}{\Delta_g}
\]
with
\[
    \Delta_g = -\pi\frac{k}{g}.
\]
The inverse Fourier transform with respect to $\theta$ of Eq.~\eqref{Phi Ideal Theta} is:  
\begin{equation}
    \hat{\phi}( x,k, g) = i\hat{\psi}_0 \arctan\left(\frac{|k|}{g}x\right).
    \label{eq:Phi Ideal x}
\end{equation}
The solution in Eq.~\eqref{eq:Phi Ideal x} just recovered can be derived also by solving Eq.~\eqref{eq:Inner Eq of Motion Theta}-\eqref{eq:Inner Induction Theta} by imposing $S=\infty$ (e.g., \cite{comisso2015extended}).  The branch cuts and poles of this solution, assuming the forcing given by the standing wave in Eq.~\eqref{eq:forcing_g}, is reported in Figure ~\ref{fig:Phi Ideal Contour}. The poles introduced by the boundary condition $\hat\Xi(k, g)$, in green, provide the long-time finite frequency response which drives the Alfv\'en resonance. The orange branch cut arises from the complex arctangent contribution and is responsible for the forced reconnection solution when $\omega_0\rightarrow0$. 

For the standing wave boundary condition, the ideal stream function can be inverted analytically by following the Bromwich contour represented in Figure ~\ref{fig:Phi Ideal Contour}. The solution in terms of the variables $x$, $y$, and $t$ is given by
\begin{equation}
    \phi = -\Phi
    \left\{\cos(\omega_0 t)\left[CI(z_+) - CI(z_-) +\ln\left(\frac{z_-}{z_+}\right)\right]
    - \frac{2 \sin(k_0 x t)}{\omega_0 t} + \sin(\omega_0 t)\left[ SI(z_+) + SI(z_-)\right]\right \},
    \label{eq:Phi Ideal t}
\end{equation}
where 
\begin{equation}
    z_{\pm} = (k_0 x\pm \omega_0) t, \quad \Phi = \frac{ \Xi_0}{\pi}\frac{\omega_0\sin[k_0 y]}{\sinh[k_0]},
\end{equation}
and $SI(z_\pm)$, $CI(z_\pm)$ are the sine and cosine integral functions, respectively. We see that for $\omega_0\neq0$ we recover the expected logarithmic singularity at the resonant site, where the phase velocity of the driven boundary perturbation, $v_{ph} = \omega_0/k_0$, is equal to the local Alfv\'en speed. The logarithmic contribution to the stream function originates from the residues at the poles $g=\pm \omega_0$, while the sine and cosine integrals arise from evaluating the jump of the Laplace integral across the branch cut. For long times, the contribution of the logarithmic response to the resonant poles dominates over that of the sine and cosine integrals, which initially provide much of the oscillatory behavior of the solution. In the limit $\omega_0\rightarrow 0$, Eq.~\eqref{eq:Phi Ideal t} describes forced reconnection at the neutral line and recovers the results of \cite{hahm1985forced}.

The ideal current density can be found  either from  $\eqref{eq:Current Density From Phi}$ in the limit of infinite $S$ or from the asymptotic expansion of \eqref{eq:Jz General z}. In either case, the ideal current density for a generic boundary condition~is 
\begin{equation}
    \Tilde{J}_z = \frac{\psi_0}{2}\frac{k}{g}\left(\frac{g}{|k|} \theta + 1\right)e^{ -\frac{g}{|k|} \theta},
     \label{eq:Current Density Ideal theta}
\end{equation}
and in the case of the standing wave boundary condition the ideal current density can be inverted to obtain
\begin{equation}
    J_z = -\Xi_0\frac{2}{\pi}\frac{k_0^4\cos(k_0 y)}{ \sinh(k_0)} \frac{x^2 t}{(k_0^2 x^2-\omega_0^2)}\left[\cos(k_0 x t)+\frac{( k_0^2 x^2-3 \omega_0^2)}{( k_0^2 x^2- \omega_0^2)} \frac{\sin(k_0 x t)}{ k_0 x t}+\frac{2 \omega_0^3}{ k_0^2 x^2 t}\frac{ \sin( \omega_0 t) }{( k_0^2 x^2-\omega_0^2)}\right].
    \label{jz_ideal_t}
\end{equation}
For $t\ll1$ Eq. \eqref{jz_ideal_t} is given by 
\begin{equation}
    J_z = - \Xi_0 \frac{4}{\pi} \frac{k_0^2 \cos(k_0 y)}{\sinh(k_0)}t\left[ 1 + \frac{1}{6}(k_0^2 x^2+\omega_0^2)t^2\right] + \mathcal{O}(t^4).
    \label{eq:jz_ideal_early_limit_t}
\end{equation}
As can be seen from Eq.~\eqref{eq:jz_ideal_early_limit_t},  the current density's amplitude increases linearly with time regardless of the forcing frequency to lowest order. As time goes by, the more general Eq.~\eqref{jz_ideal_t} recovers the result of \cite{hahm1985forced} for $\omega_0=0$, where the amplitude of $J_z$ increases linearly with time at $x=0$, and it is localized in a region whose width decreases as $1/t$. Outside the origin, the current density's amplitude still increases linearly with time but its spatial behavior is highly oscillatory. For $\omega_0\neq0$, the current density at the origin is a purely oscillating function. For long times, however, it grows faster than linear near the resonant point where it builds up by eventually leading to a singularity.

\section{The Resistive Regime}
We separate now the discussion of the time-asymptotic  behavior, corresponding to $Q\ll1$, into two cases: the case of small or zero frequency, $\omega_0\ll S^{-1/3}$, for which we expect to describe forced magnetic reconnection, and the case of high frequency $\omega_0\gg S^{-1/3}$. 

\subsection{Forced Reconnection}
In the limit $Q \ll 1$ (or $g \ll S^{-1/3}k^{2/3}$) Eq.~\eqref{eq:Phi Inner} with Eq. \eqref{eq:A} approximates to 
\begin{equation}
    \Tilde{\phi} = -\frac{\Gamma[\frac{3}{4}]}{2 \pi} \psi_0 \frac{g}{k} \left( \frac{1}{|k|}\sqrt{\frac{g}{S}}\theta^2\right)^{1/4} K_{\frac{1}{4}}\left(\frac{1}{|k|}\sqrt{\frac{g}{S}} \frac{\theta^2}{2}\right),
    \label{eq:Phi Const Theta}
\end{equation}
where $K_\nu$ is the modified Bessel function of the second kind. This is colloquially known as the constant-psi regime, or phase D of \cite{hahm1985forced}. In this limit, both $\Delta_g$ and $\Delta_0$ are retained in $\hat{\psi}_0$ with $\Delta_g$ approximated~to 
\begin{equation}
    \Delta_g = \sqrt{2} (\frac{g^5 S^3}{k^2})^{1/4} \Gamma(3/4)^2.
\end{equation}
From the definition of $\Delta_g$, we observe the time scale $t \sim S^{3/5}$ after which reconnection reaches a steady state \cite{hahm1985forced,cole2004forced}. The flux function at the origin $\hat\psi_0$, given below, is consistent with that found by \cite{hahm1985forced,cole2004forced}, albeit  for different boundary conditions,
\begin{eqnarray}
        \hat{\psi}_0 &=& \frac{\frac{2k}{\sinh(k)}\hat{\Xi}}{\sqrt{2} (\frac{g^5 S^3}{k^2})^{1/4} \Gamma(3/4)^2 + 2\frac{k}{\tanh(k)}}.
        \label{eq:Psi_0 Small Q}
\end{eqnarray}
\begin{figure}
    \centering
    \includegraphics[width=.4\textwidth]{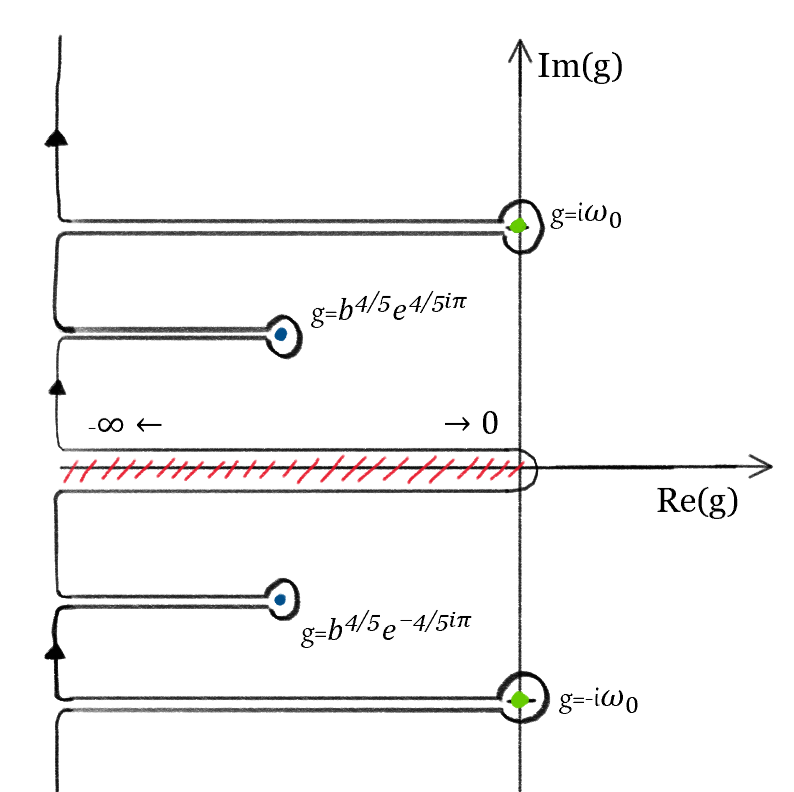}
    \caption{The branches and poles of the constant-psi solution \eqref{eq:Phi Const g} are shown here in the complex plane. In green are the resonant poles of the boundary condition, in blue the singular contributions of the $g^{5/4}$ root, and in red is the branch of the Modified Bessel and Struve functions which extends from $g=-\infty$  to $g=0$.}
    \label{fig:Reconnection Contour}
\end{figure}

The stream function in Eq. \eqref{eq:Phi Const Theta} can be inverted from Fourier space to configuration space yielding, for the standing wave boundary condition, the following result, 

\begin{equation}
    \hat{\phi} = \Xi_0\sqrt{\frac{\pi}{2}}\Gamma(\frac{3}{4})\frac{\sin(k_0  y)}{\sinh(k_0)}\frac{g^2}{(g^2+\omega_0^2)}\left(k_0\sqrt{\frac{S}{g}}\right)^{3/4} 
    \sqrt{x}\left(
    \frac{ I_{1/4}\left(k_0\sqrt{\frac{S}{g}} \frac{x^2}{2} \right) - L_{1/4}\left( k_0\sqrt{\frac{S}{g}} \frac{x^2}{2} \right)}{ \sqrt{2}(\frac{g^5 S^3}{k_0^2} )^{1/4}\Gamma(3/4)^2 + 2\frac{k_0}{\tanh(k_0)}}\right),
    \label{eq:Phi Const g}
\end{equation}
where $I_\nu$ is the modified Bessel function of the first kind and $L_\nu$ is the modified Struve function. The branch and poles of this solution with the Bromwich contour are sketched in Figure \ref{fig:Reconnection Contour}. The green poles represent once more the resonant contributions of the forcing that merge with the branch cut, represented in red, in the limit of $\omega_0\rightarrow0$. The blue poles are the roots of the denominator in the last term in Eq. \eqref{eq:Phi Const g}. It is noted that we recover the same branches and poles as \cite{hahm1985forced}, but in their case the resonant poles of the boundary overlap with the branch cut at $g=0$.

To find the time dependent stream function in this regime, we use Mathematica to numerically take the inverse Laplace transform of Eq. \eqref{eq:Phi Const g} about the Bromwich contour just discussed. In the limit $\omega_0\rightarrow0$, the contributions to the integral are the residues of Eq. \eqref{eq:Phi Const g} evaluated at the $g^{5/4}$ poles (the blue poles in Fig.~\ref{fig:Reconnection Contour}) and the jump across the branch extending from $g=-\infty$ to $g=0$. For small but non zero frequencies $\omega_0$, we must include also the contribution of the residues at $g=\pm i\omega_0$, which add an oscillatory contribution to the solution that is non-negligible for $t\rightarrow\infty$. 

\begin{figure}
    \centering
    \includegraphics[width=.95\textwidth]{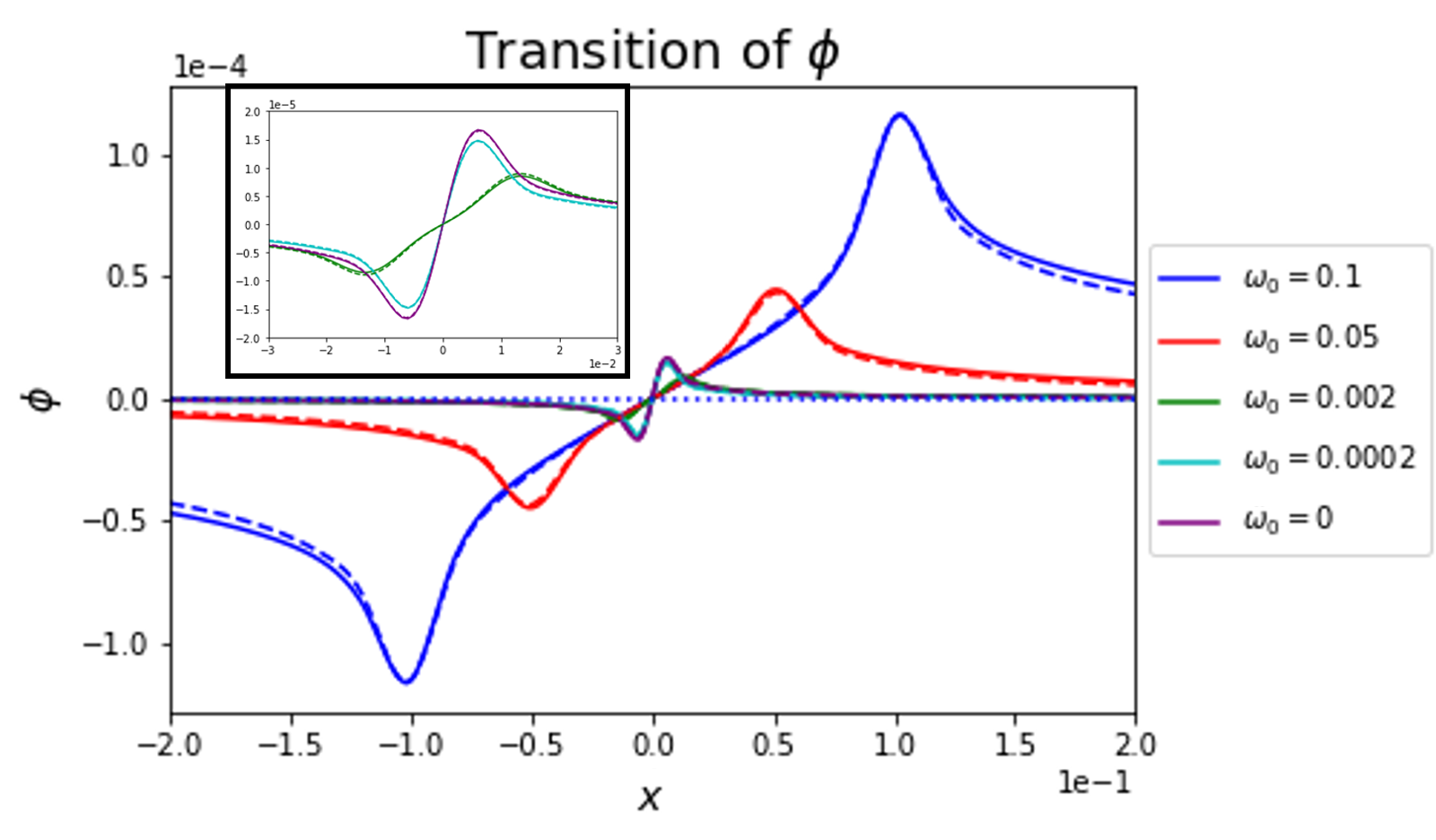}
    \caption{The time-asymptotic solution for the stream function $\phi (x,y,t)$  (dashed lines) compared with numerical simulations (solid lines). The dashed green, cyan, and purple lines are the inverse Laplace transformed constant-psi solutions, Eq. \eqref{eq:Phi Const g}, for $\omega_0 = 0.002$, $\omega_0 = 0.0002$, and $\omega_0 = 0$, respectively. These solutions are best seen in the inset plot, showing an enlarged picture of the region around the neutral line. The dashed blue and red lines are the inverse Laplace transformed Alfv\'en resonance solutions, Eq. \eqref{eq:Phi Resistive Theta}, for $\omega_0 = 0.1$ and $\omega_0 = 0.05$, respectively.
    The constant-psi solution for $\omega_0 = 0.1$  has been also plotted (dotted line) to demonstrate the breakdown of the constant-psi solution as the system transitions towards Alfv\'en resonance for increasing frequencies. }
    \label{fig:Phi Transition}
\end{figure}

The resulting stream function $\phi( x, y, t)$ is compared with results from a numerical simulation in Figure \ref{fig:Phi Transition}, where dashed lines correspond to the theoretical solutions and solid lines correspond to results from ab-initio numerical simulations. Details on the numerical code can be found in the Appendix. The  constant-psi solution is represented by the purple, cyan and green colors for $\omega_0=0$, $\omega_0=0.0002$,  and $\omega_0=0.002$, respectively, where we have fixed $y=\pi/2$ and $t = 3000$, which is sufficiently long compared to the resistive time scale $t \sim S^{1/3}$. The chosen frequencies all satisfy the condition $\omega_0\ll S^{-1/3}$ and, as can be seen, the solution corresponding to the constant-psi regime matches very well with the simulation results for each frequency.

The current density is found by taking the inverse Fourier transform of \eqref{eq:Current Density From Phi} in the limit \textcolor{blue}{$Q \ll \theta^2$}

\begin{equation}
    \hat{J}_z = -(Sk)\int^{\infty}_{-\infty}e^{ix\theta}\frac{d}{d\theta}\Tilde{\phi},
    \label{eq:Jz integral phi}
\end{equation}
which is identically
\begin{equation}
    \hat{J}_z = - i(Sk)x\hat{\phi}(x,k,g),
    \label{eq:Jz hat x}
\end{equation}
from the definition of the Fourier transform. By using the appropriate $\hat{\phi}$ for this limit for constant-psi (from equation \eqref{eq:Phi Const g} in Fourier $k$ space) $\hat{J_z}$ is given by
\begin{equation}
    \hat{J}_z = \Xi_0\sqrt{\frac{\pi}{2}}\Gamma(\frac{3}{4})\frac{k_0 \cos(k_0 y)}{\sinh(k_0)}\frac{g^2 S}{(g^2 + \omega_0^2)}\left(k_0\sqrt{\frac{S}{g}} x^2\right)^{3/4}  \left(\frac{I_{1/4}\left(k_0\sqrt{\frac{S}{g}}\frac{x^2}{2}\right) - L_{1/4}\left(k_0\sqrt{\frac{S}{g}}\frac{x^2}{2}\right)}{\sqrt{2}(\frac{ g^5 S^3}{k_0^2} )^{1/4}\Gamma(3/4)^2 + 2\frac{k_0}{\tanh(k_0)}}\right).
    \label{eq:Jz Constant g}
\end{equation}
We note that the current density has the same branches and poles as the stream function in Eq.~$\eqref{eq:Phi Const g}$, outlined in Figure \ref{fig:Reconnection Contour}. The current $J_z$ can thus be found as a  function of time by computing the inverse Laplace transform over the same contour as we did for $\phi$.

In principle, Eq. \eqref{eq:Jz Constant g} describes also the current density at the origin under an appropriate limit by expanding the Modified Bessel and Struve functions, $I_{1/4}$ and $L_{1/4}$ respectively, for small arguments. For simplicity, to find the current density at $x=0$ required to calculate the reconnection rate, we follow the method adopted in other works by employing the relation
\begin{eqnarray}
    J_z(0, y, t) = S \frac{\partial \psi_0}{\partial t} = S \mathcal{L}^{-1} (g\hat{\psi}_0),
    \label{eq: Jz at 0}
\end{eqnarray}
obtainable from \eqref{eq:Jz integral phi} by imposing $x=0$ in the exponential,  making use of the relationship of $\tilde{\phi}$ and $\tilde\psi$, Eq. \eqref{eq:Inner Induction}, and, finally, by applying the limit  \textcolor{blue}{$Q \ll \theta^2$}. We have also assumed here that reconnection has been developed sufficiently such that $\psi$ is approximately constant across the layer, as is the case for the constant-psi approximation. For the boundary condition employed in this paper, Eq.~\eqref{eq: Jz at 0} yields,
\begin{equation}
    \hat{J}_z(0, y, g) = -\Xi_0\frac{2 k_0\cos(k_0 y)}{\sinh(k_0)}\frac{g^2S}{(g^2 + \omega_0^2)}
    \left(\frac{1}{\sqrt{2} (\frac{g^5 S^3}{k_0^2})^{1/4} \Gamma(3/4)^2 + 2\frac{k_0}{\tanh(k_0)}}\right).
    \label{eq:jz 0 Const g}
\end{equation}
We now evaluate numerically its inverse Laplace transform via the Bromwich contour in Figure \ref{fig:Reconnection Contour}  (using Mathematica) to calculate the reconnection rate $R(t) = J_z(0,t)/S$ at a given position $y$.

\begin{figure}
    \centering
    \includegraphics[width=.3\textwidth]{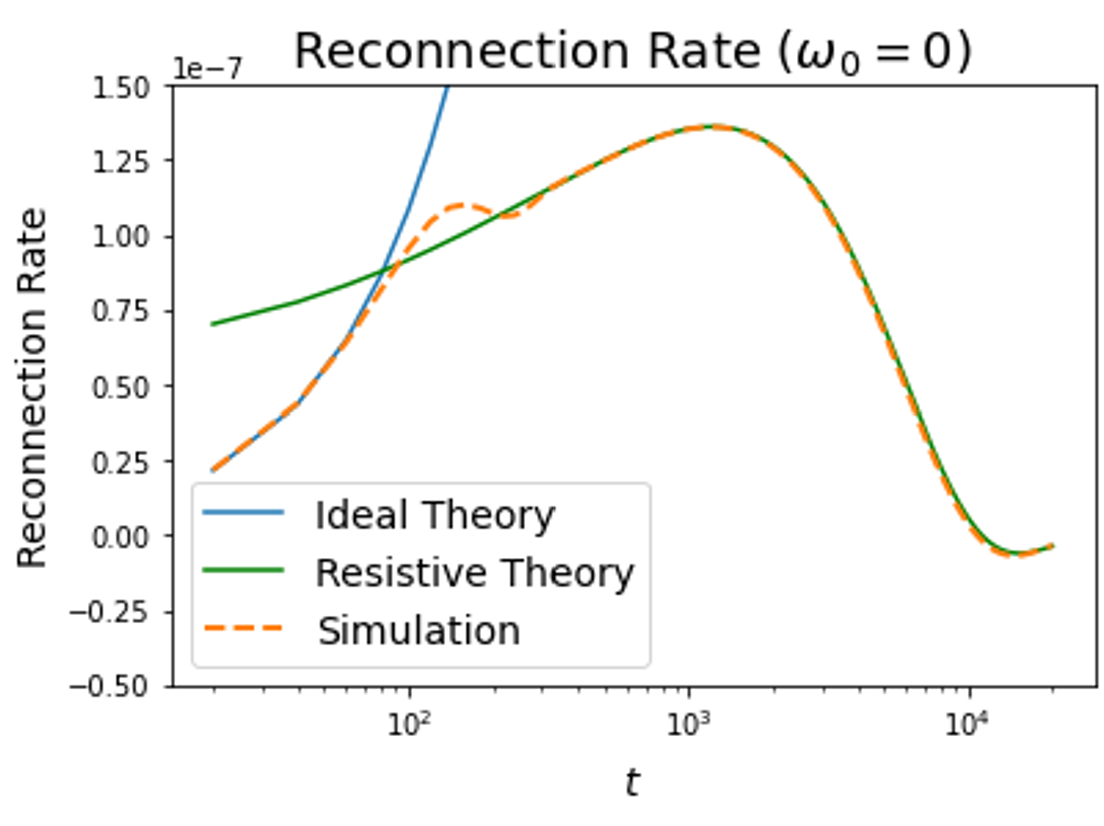}
    \includegraphics[width=.3\textwidth]{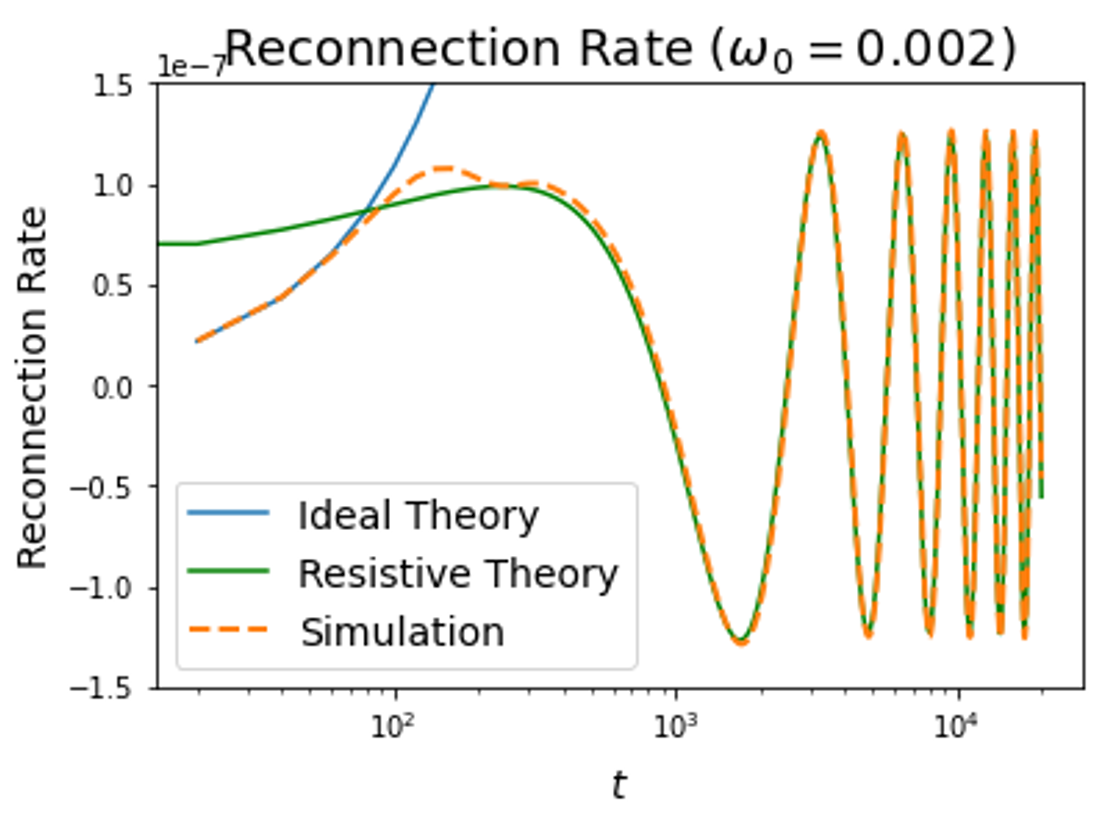}
    \includegraphics[width=.31\textwidth]{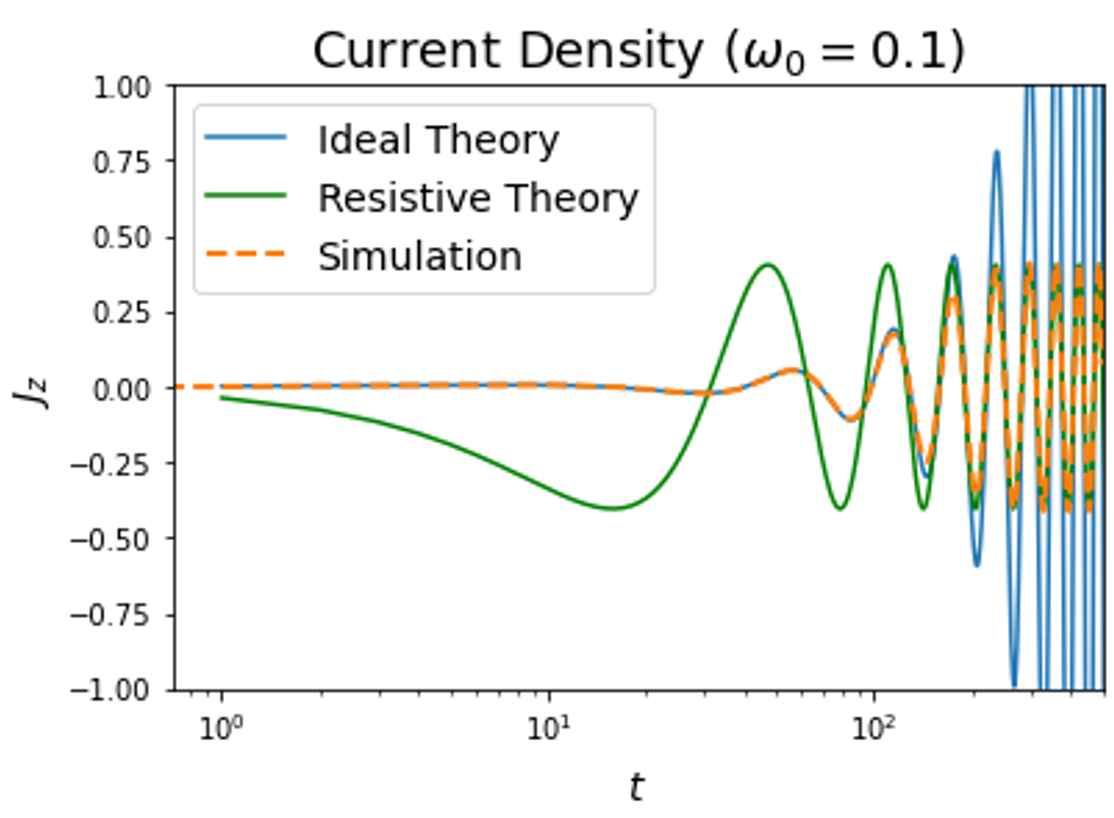}
    \caption{\label{fig:Jz Time Comparison} 
    The left and middle panels show the reconnection rate as a function of time from  numerical simulations compared with the two asymptotic solutions: the ideal solution, Eq. \eqref{jz_ideal_t} divided by $S$ in blue, and the inverse Laplace transform of the constant-psi solution, Eq. \eqref{eq: Jz at 0} in green. We show results for $\omega_0 = 0$ (left panel) and $\omega_0 = 0.002$ (middle panel). In the right panel, the simulated current density for $\omega_0 = 0.1$ has been plotted at the resonant site. It is compared with the  same ideal solution (blue color) and with the  resistive solution solution given by the  inverse Fourier transform of Eq. \eqref{eq:Current Density Resistive theta} (green color). The parameters used here are $S = 10^6$, $k_0 = 1$, $y = \pi/2$.}
\end{figure}

The time evolution of the reconnection rate $R(t)$ in the cases of $\omega_0=0$ and $\omega_0=0.002$ is plotted in Figure \ref{fig:Jz Time Comparison}, left and middle panels, where we compare $R$ from the 2D simulation (orange dashed line) with the ideal (blue) and constant-psi (green) solutions. The reconnection rate behaves as expected for the zero frequency case, (left panel): it initially grows linearly with time as described by the ideal solution and for $t\gg S^{1/3}$ it follows the constant-psi solution. After times much longer than the reconnection time scale $t \sim S^{3/5}$, the reconnection rate approaches zero as the flux function at the origin reaches a steady state. The reconnection rate has a first peak in the intermediate regime, between the ideal and constant-psi regimes, when the resistive layer forms at $t\sim S^{1/3}$.   In the case of a small but non zero frequency (middle panel), the reconnection rate follows an analogous evolution but, over a longer time scale, instead of reaching a steady state the reconnection rate oscillates in response to the boundary driver. 
The fact that the small but non zero frequency case also gives rise to reconnection isn't surprising, as discussed in the Introduction, and was analyzed previously by \cite{luan2014transition} for the eigenvalue problem and by \cite{fitzpatrick1991interaction} in the study of the interaction of resonant magnetic perturbations with rotating plasmas. In a later section we discuss in more details the effects of finite frequency on reconnection. 

\subsection{Transition to Alfv\'en Resonance}

In the previous subsection, we analyzed the constant-psi regime of Eq. \eqref{eq:Phi Inner} for small frequencies ($\omega_0 \ll S^{-1/3}$). As discussed, when $\omega_0\ll S^{-1/3}$ the solution corresponding to the constant-psi regime matches very well with the simulation results. However, as the frequency is increased beyond $\omega_0\sim S^{-1/3}$, the agreement diverges and the constant-psi solution  breaks down. This is shown for example by the dotted blue line in Figure \ref{fig:Phi Transition}, that represents the constant-psi solution but for a high frequency case, $\omega_0 = 0.1$; as can be seen, the dotted line does not reproduce simulation results, which are represented by the solid blue line. Therefore, for large frequencies we must consider a different limit to describe  Alfv\'en resonance. 

For large frequencies, $\omega_0 \gg S^{-1/3}$, the residues at the resonant poles of the boundary driver, represented in green in Figure \ref{fig:Reconnection Contour}, contribute to the time-asymptotic response of the system. The contributions from both the jump at the branch cut and the residues at the poles located in the region $\Re({g})<0$ have vanishing contributions for $t\rightarrow\infty$. Additionally, if the frequency is large enough such that $\omega_0\gg S^{-1/3}k^{2/3}$, one cannot use the asymptotic expansion of Eq. \eqref{eq:Phi Inner} and \eqref{eq:A} for $Q\ll 1$ to evaluate the residues at $g=\pm i \omega_0$, as it was done for small frequency. Instead,  one should use the asymptotic expansion for $Q\gg1$ to calculate the residues at those poles, except this time by retaining resistive corrections.  As gradients become strong around  the resonant point, the effects of resistivity will be to resolve and smooth out the logarithmic singularity of the ideal solution. 
Performing the same expansion for $Q\gg 1$ as with the ideal case but by retaining higher-order terms in $z/Q^{3/2}\ll 1$, we find that the stream function can be approximated by
\begin{eqnarray}
    \Tilde{\phi}(\theta,y,g) = \frac{\hat{\psi}_0}{2}\frac{1}{\theta}e^{-\frac{g}{|k|} \theta\left(1 + \frac{1}{6}\frac{\theta^2}{gS}\right)+\mathcal{O}\left(\frac{\theta^5}{g S^2}\right)}\left[  1+\mathcal{O}\left( \frac{\theta^2}{g S} \right)\right],
    \label{eq:Phi Resistive Theta}
\end{eqnarray}
which recovers the ideal solution in the limit of infinite $S$, as well as the same exponential dependence found in \cite{bertin1986alfven} in the context of  normal mode analysis. As just discussed, in the time asymptotic limit, the dominant contribution is given by the residues evaluated at the resonant poles of the boundary condition, which in our particular case give 
\begin{equation}
    \Tilde{\phi}(\theta,y,t) = i \frac{\Xi_0 }{\pi}\frac{\omega_0\sin(k_0 y)}{\sinh(k_0)}\sin\left(\omega_0(t-\frac{\theta}{k_0})\right)
    \frac{e^{ - \frac{1}{6} \frac{\theta^3}{Sk_0}}}{\theta}.
    \label{eq:Phi Resistive t}
\end{equation}

To find the spatio-temporal stream function we finally used Mathematica to evaluate the inverse Fourier transform from $\theta$ to $x$ of Eq.~\eqref{eq:Phi Resistive t}. The resulting solution for $\omega_0=0.1$ and $\omega_0=0.05$ is reported in Figure \ref{fig:Phi Transition}, dashed blue and red lines, and compared with simulation results represented by the solid blue and red lines. It is apparent that Eq.~\eqref{eq:Phi Resistive Theta}, when Fourier-inverted,  captures the long-time behavior of Alfv\'en resonances. We see that the contribution of the resistivity is both to remove the logarithmic singularity and to provide a finite width to the resonant boundary layer. The boundary layer, as can be seen by inspection of Eq.~\eqref{eq:Phi Resistive t}, scales as $\delta\sim S^{-1/3}$.

The current density is evaluated from the expansion of \eqref{eq:Jz General z} while maintaining higher orders in the exponential, 
\begin{equation}
     \Tilde{J}_z(\theta,k,g) = \frac{\psi_0}{2}\frac{k}{g}\left(\frac{g}{|k|} \theta + 1\right)e^{ -\frac{g}{|k|} \theta\left(1 + \frac{1}{6}\frac{\theta^2}{gS}\right)}.
     \label{eq:Current Density Resistive t}
\end{equation}
Again we evaluate the residues at the resonant poles of the boundary condition to find the long time behavior, yielding the following solution to be numerically inverted from $\theta$ to configuration space, 
\begin{equation}
    \Tilde{J}_z(\theta,y,t) = -\frac{\Xi_0}{\pi}\frac{k_0 \cos(k_0 y)}{\sinh(k_0)}e^{ - \frac{1}{6} \frac{\theta^3}{Sk_0}}\left( \cos(\zeta) - \sin(\zeta) \frac{\omega_0}{k_0}\theta \right),
    \label{eq:Current Density Resistive theta}
\end{equation}
where $\zeta = \omega_0(t-\frac{\theta}{k_0})$. The current density  at the resonant site, found through numerical inversion of \eqref{eq:Current Density Resistive theta}, has been plotted in the right frame of Figure \ref{fig:Jz Time Comparison}. Just like with the reconnection rate for small frequencies, we see very good agreement between the ideal solution and the numerical simulation up to nearly $t\sim S^{1/3}$. Over longer time scales, the system reaches a steady-state due to resistive dissipation, and the solution in Eq.~\eqref{eq:Current Density Resistive theta} provides the correct time-asymptotic behavior where an oscillating steady-state is reached.

Figures \ref{fig:Phi Transition} and \ref{fig:Jz Time Comparison} together demonstrate the transition from Alfv\'en resonance to forced magnetic reconnection as $\omega_0$ varies from $\omega_0\gg S^{1/3}$ to $\omega_0\ll S^{1/3}$. As the frequency approaches zero, the resonant sites shifts towards the neutral line, since Alfv\'en resonance necessarily occurs where the phase velocity of the driver is equal to the local Alfv\'en speed, $\omega_0/k=v_a(x)$.  When the resonant layers, which are located on either side of the neutral line, overlap across the neutral line as a consequence of their finite widths, the resonance couples with the reconnection mode. The subsequent section discusses the effect of finite frequency on such a coupling.

\subsection{Discussion: scalings laws and effect of finite frequency on reconnection}

\begin{figure}
    \centering
    \includegraphics[width=.45\textwidth]{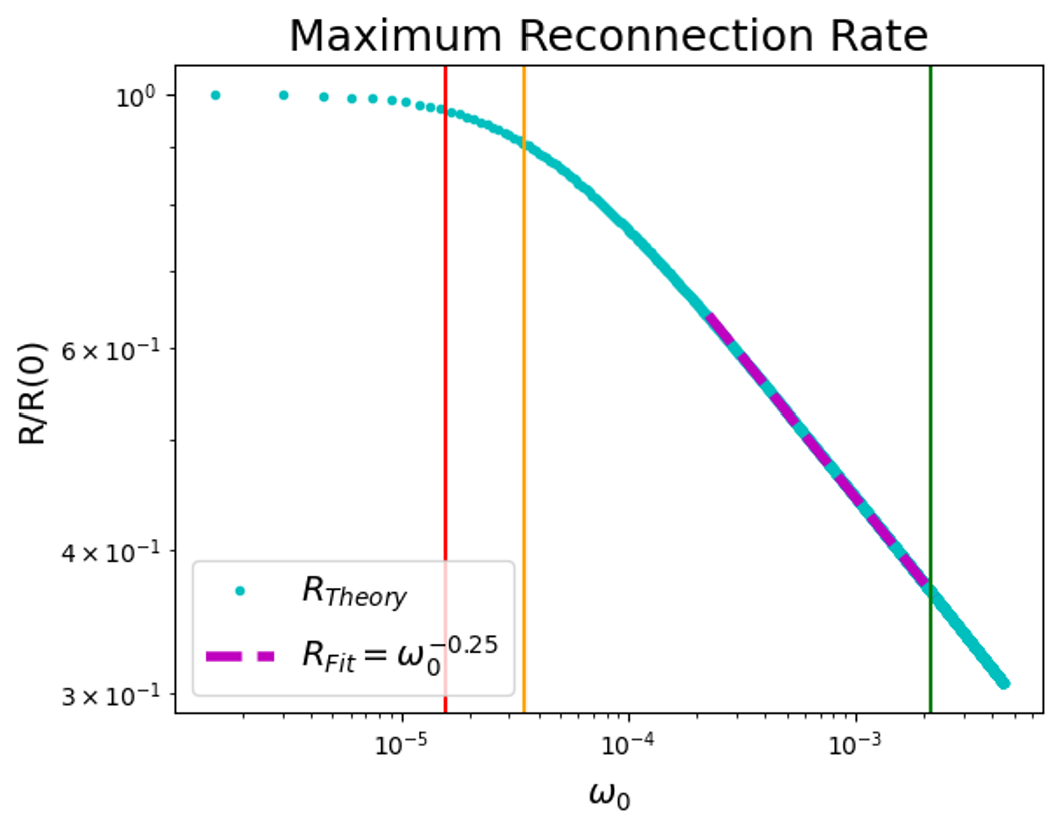}\includegraphics[width=.43\textwidth]{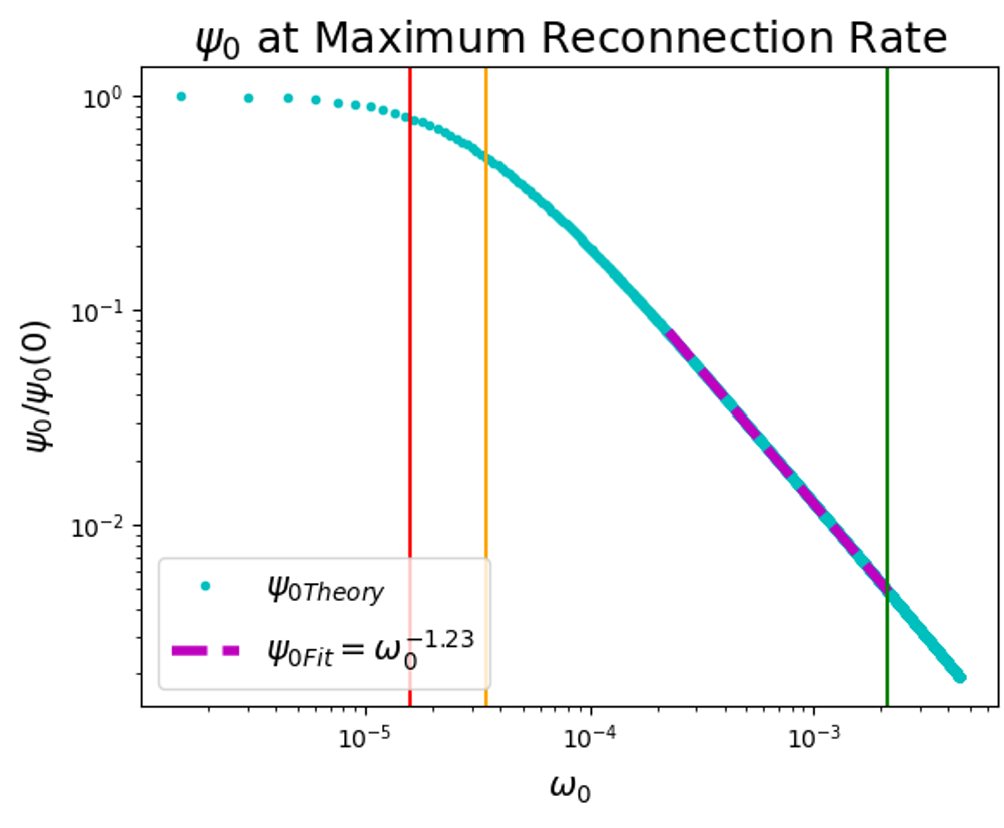}
    \caption{Plots of the reconnection rate and reconnected flux as functions of $\omega_0$  evaluated at the time of maximum reconnection rate in the constant-psi regime, for $S=10^8$. They are calculated from the inverse Laplace transforms of Eq. \eqref{eq: Jz at 0}, divided by $S$, and Eq. \eqref{eq:Psi_0 Small Q}, which are applicable for $\omega_0 \ll S^{-1/3}$. The red (left most) line in each plot indicates $\omega_0 = S^{-3/5}$. The orange (middle) line indicates the location of half maximum half width of $\psi_0$. The green (right most) line indicates $\omega_0 = S^{-1/3}.$ In the frequency range beyond the half-width of $\psi_0$ but below $\omega_0 = S^{-1/3}$, we observe a power law scaling  $R\sim\omega_0^{-0.25}$ and $\psi_0\sim\omega_0^{-1.23}$.}
    \label{fig:Omega Scalings}
\end{figure}

To further understand the effect of finite frequency on forced reconnection, in this section we analyze two sets of scalings predicted by the theory. The first set is the scaling of the width $\delta$ of the boundary layer for forced reconnection and Alfv\'en resonance as functions of the Lundquist number $S$. The second set is the scaling of the constant-psi reconnection rate and reconnected flux at the neutral line as functions of the driving frequency $\omega_0$.  

Through inspection of the current density equation in the constant-psi regime which describes forced reconnection, Eq. \eqref{eq:Jz Constant g}, we find the theoretical scaling law for $\delta$. For this regime we assume the scaling $g \sim S^{-3/5}$, found from the poles in the denominator. Upon substitution of $
g$ for $S$ in Eq. \eqref{eq:Jz Constant g} we find the argument of the Modified Bessel and Struve functions to be of order unity  when $x \sim S^{-2/5}$. This indicates a layer width scaling  $\delta \sim S^{-2/5}$, which is the known  scaling for this process \cite{hahm1985forced}. 

For Alfv\'en resonance we perform the same analysis of its equivalent current density equation, applicable for large driving frequencies $\omega_0 \gg S^{-1/3}$, Eq. \eqref{eq:Current Density Resistive theta}. For this scaling we need only the argument of the exponential to determine the inner layer width dependence on $S$. We see that the argument is of order unity when $\theta \sim S^{1/3}$ which is indicative of a layer width scaling $\delta \sim S^{-1/3}$, as expected \cite{mok1985resistive,kappraff1977resistive,bertin1986alfven}. 

Although it is not possible to provide a theoretical solution for the intermediate regime when $Q\sim 1$ (which characterizes the region at which our approximations mismatch with simulation in Figure \ref{fig:Jz Time Comparison}), we can nevertheless use the general solution given in Eq.~\eqref{eq:Phi Inner} to infer the expected $\delta$. From the definition of the parameter $Q$, Eq. \eqref{eq:Q}, the scaling $g \sim S^{-1/3}$ naturally emerges. Replacing $g\sim S^{-1/3}$ in Eq.~\eqref{eq:Phi Inner} we find the argument of the Tricomi confluent hypergeometric functions of Eq.~\eqref{eq:Phi Inner}, $z = \sqrt{Q}\theta/(S k)^{2/3}$, is of order 1 for $\theta \sim S^{1/3}$ or $\delta \sim S^{-1/3}$. The regime $Q \sim 1$, which corresponds to $t \sim S^{1/3}$, applies to both Alfv\'en resonance and forced reconnection. In particular, it shows that forced reconnection  starts off in the non-constant-psi regime, as first discussed in \cite{hahm1985forced}, before continuing on through the constant-psi. Thus, we see that when resistivity becomes important at $t\sim S^{1/3}$ both forced reconnection and Alfv\'en resonance exhibit the same layer width, allowing for their coupling, when the forcing frequency is sufficiently small.

We make now analytical predictions of how the reconnection rate and the reconnected flux scale with the forcing frequency $\omega_0$. We conduct this analysis under the constant-psi regime, i.e., $R$ is calculated by using Eq. \eqref{eq: Jz at 0} divided by $S$, and \eqref{eq:Psi_0 Small Q} is used for $\psi_0$, that in this approximation represents the reconnected flux at $x=0$. 

For frequencies  $\omega_0\ll S^{-3/5}$, the system has time to evolve from the ideal to the constant-psi regime almost unaffected by the oscillating driver, and we expect that for such low frequencies both $R$ and $\psi_0$ are unaffected by $\omega_0$. When the frequency approaches the inverse of the constant-psi time scale, which is around the time when the peak in reconnection occurs (see Fig.~\ref{fig:Jz Time Comparison}), we expect the reconnection process to be affected by the finite frequency driver.   Now, in the frequency range $S^{-3/5}\ll \omega_0 \ll S^{-1/3}$ one can neglect the term $\Delta_0 = 2 k_0/\tanh(k_0)$ in the denominator of Eqs. \eqref{eq:Psi_0 Small Q} and \eqref{eq: Jz at 0}. Additionally, as demonstrated previously, as $\omega_0$ increases the contribution to the long time solution is provided by residues at $g=\pm i\omega_0$. By analyzing under this light the residues of the inverse Laplace transforms of Eqs. \eqref{eq: Jz at 0} and \eqref{eq:Psi_0 Small Q} at $g = \pm i \omega_0$, it can be shown that the expected scalings are $R \sim \omega_0^{-1/4}$ and $\psi_0 \sim \omega_0^{-5/4}$. This would suggest that although $\omega_0\sim S^{-1/3}$ determines the transition from forced reconnection to Alfv\'en resonance, the reconnection rate is reduced in the range of frequencies $S^{-3/5}\ll \omega_0 \ll S^{-1/3}$. For frequencies larger than $\omega_0\gg S^{-1/3}$ reconnection is completely suppressed by the Alfv\'en resonance.

To measure the effect of finite frequency on reconnection, in Figure \ref{fig:Omega Scalings} we show  the plot of the normalized, maximum reconnection rate $R$ and corresponding reconnected flux function $\psi_0$ as functions of $\omega_0$ in the left and right panel, respectively, for $S=10^8$. As explained above, the constant-psi regime was assumed in making these plots. Therefore, we used Eqs. \eqref{eq: Jz at 0}, divided by $S$, for $R$ and \eqref{eq:Psi_0 Small Q} for $\psi_0$, and calculated their inverse Laplace transforms at the time for which the reconnection rate is at its maximum. With the vertical red, orange and green (left, central, and right) lines, we mark $\omega_0 = S^{-3/5}$, the half-width of $\psi_0$, and  $ \omega_0 = S^{-1/3}$. As can be seen, the predicted scalings of $R \sim \omega_0^{-1/4}$ and $\psi_0 \sim \omega_0^{-5/4}$ are very closely matched in Figure \ref{fig:Omega Scalings}, in which we found power law scalings of $R\sim\omega_0^{-0.25}$ and $\psi_0\sim\omega_0^{-1.23}$ for the reconnection rate and reconnected flux, respectively, shown by the fits with the dashed line. That is we find, as expected, a strong suppressing effect on reconnection for frequencies less than the transition frequency $\omega_0 \sim S^{-1/3}$ and higher than $\omega_0\sim S^{-3/5}$.

\section{Summary and conclusions}

In this paper we have derived and analyzed the general solution to the resistive, linearized MHD equations (Eqs. \ref{eq:Scalar Eq of Motion}-\ref{eq:Scalar Induction}) describing the evolution in time of a plasma with a sheared magnetic field (Eq. \ref{eq:Background Mag}) subjected to an oscillatory boundary perturbation turned on at time $t=0$ (Eqs. \ref{eq:Scalar psi BC}-\ref{eq:Scalar phi BC}). Through  asymptotic matching, we determined the spatial dependence (Eq. \eqref{eq:Phi Inner}) and the amplitude (Eq. \eqref{eq:A}) of the general solution for the perturbed stream function in the inner layer. Such a solution shows explicitly that the key parameter to describe different regimes in the evolution of the system, as well as the transition from forced reconnection to Alfv\'en resonance, is $Q=g S^{1/3} k^{-2/3}$. We have then discussed two asymptotic limits of Eq. \eqref{eq:Phi Inner}, $Q\gg1$ and $Q\ll1$, and we demonstrated that our theory describes  Alfv\'en  resonance when the frequency of the forcing is $\omega_0 \gg S^{-1/3}$, and forced reconnection for small to zero frequency, $\omega_0 \ll S^{-1/3}$. Spatio-temporal solutions have been derived for a standing wave boundary condition, given in Eq. \eqref{eq:forcing}, which we compared with linear simulations.  

In the ideal limit, defined by $Q\gg1$ and $z/Q^{3/2}\ll1$ (up to first order), our general solution recovers that of forced reconnection for $\omega_0 = 0$; for $\omega_0 \neq 0$, our solution describes Alfv\'en resonance. In particular, the known logarithmic singularity at the resonance point where $\omega_0/k=v_a(x)$ is due to the residues at the poles of the boundary driver, and contribute to the time-asymptotic response of the system, as can be seen from the Bromwich contour in Figure \ref{fig:Phi Ideal Contour}.  

The non-ideal time-asymptotic response of the system corresponds to $Q\ll1$. We have provided the corresponding asymptotic expansion of the inner layer stream function for $Q\ll1$, by recovering the constant-psi solution that describes forced  reconnection for times $t\gg S^{1/3}$.
The corresponding  Bromwich contour is shown in Figure \ref{fig:Reconnection Contour}. 

For large frequencies, the time-asymptotic response is dominated by the residues at the forcing frequency (see Fig. \ref{fig:Reconnection Contour}), and if the condition $\omega_0 S^{1/3}k^{-2/3}\gg1$ is satisfied, then a different expansion of the stream function for their evaluation is required. In this case, the solution is given by the asymptotic expansion of the inner layer solution for $Q\gg1$ with terms in $z/Q^{3/2}\ll 1$ retained to higher order. This regime  represents Alfv\'en resonance in the time-asymptotic limit.

A comparison of the theoretical predictions and numerical simulations is given in Figure \ref{fig:Phi Transition} and \ref{fig:Jz Time Comparison}, testing our model's ability to capture both forced reconnection and Alfv\'en resonance.  We show that for large frequencies the resonant poles of the boundary driver were sufficiently far from zero (accordingly, the resonant points are far from the neutral point) and a pair of anti-symmetric resonances, of width $\delta \sim S^{-1/3}$, are clearly defined in that case. As these poles are brought towards $\omega_0=0$ (the resonant points approach the neutral point $x=0$) in the small frequency limit, the widths of the resonant layers overlap across $x=0$ by coupling to the reconnection mode. The rate of reconnection and its signatures are strongest when the poles overlap entirely, that is for $\omega_0 = 0$. 

Finally, we demonstrated that our solution recovers known scalings of the resonant layer $\delta$ as a function of the Lundquist number $S$ under specific limits. Additionally, we demonstrated new scalings of the reconnection rate and reconnected flux as functions of the driving frequency of the boundary condition. In Figure \ref{fig:Omega Scalings} we show that, in the constant-psi regime, the maximum reconnection rate and reconnected flux are largest for frequencies less than $\omega_0 \lesssim S^{-3/5}$ and is reduced at larger frequencies until reconnection is completely suppressed for  $\omega_0 \gg S^{-1/3}$. 

We conclude that forced reconnection is indeed a particular case of Alfv\'en resonant absorption and there is a transition from Alfv\'en resonance to forced reconnection when the forcing frequency $\omega_0\ll S^{-1/3}$. However, the reconnection rate and reconnected flux are strongly reduced for $\omega_0\gg S^{-3/5}$.  When the forcing frequency is $\omega_0\gg S^{-1/3}$, Alfv\'en resonance decouples from the reconnecting mode and the formation of resonant layers away from the neutral line effectively shields reconnection.

Here we have  studied the interaction between Alfv\'en resonances and forced reconnection by considering an idealized symmetric forcing away from the neutral line. However, our approach is general and provides the first step for further investigations of resonances in diverse environments and within different plasma models. In particular, more realistic perturbations can be considered, such as asymmetric and impulsive. This type of forcing is especially of interest for reconnection onset at Earth's magnetopause. In this regard, recent observations from the MMS spacecraft  within Earth's magnetosphere \cite{burch2016electron} have identified reconnection events taking place at electron scales, termed electron-only reconnection \cite{phan2018electron,huang2021electron}. By building on prior investigations into forced reconnection within the EMHD framework \cite{avinash1998forced}, our approach can be extended to investigate the effect of a time-dependent forcing on electron-only reconnection, with applications to reconnection onset within Earth's magnetosphere. Additionally, our unified approach holds potential implications for studying the transition from wave-based to reconnection-based coronal heating scenarios \cite{velli2015models,malara1994wave}, which will be explored in future work.

\begin{acknowledgments}
This research was supported by NSF grant 2108320. One of us (AT) was also supported by NSF CAREER grant 2141564 and another (FW) by DOE grant DE-FG02-04ER54742. We also acknowledge the Texas Advanced Computing Center (TACC) at The University of Texas at Austin for providing HPC resources that have contributed to the research results reported within this paper. 
\end{acknowledgments}

\appendix

\section{Numerical Simulations}
The numerical code used to validate our analytical solutions is a 2D incompressible MHD code. The normalized equations that we integrated are: 

\begin{equation}
    \frac{\partial \psi}{\partial t} = x\frac{d\phi}{d y} + \frac{1}{S} \nabla^2\psi 
\end{equation} 
\begin{equation}
    \frac{\partial \Omega}{\partial t} = -x\frac{d}{d y}\nabla^2\psi 
\end{equation}
where $\Omega=-\nabla^2\phi$ is the vorticity. We impose a boundary condition at $ x=\pm 1$ with small amplitude given by:
\begin{equation}
    \psi( x=\pm 1, y,  t) = \Xi_0 \cos( k_0 y)\cos(\omega_0  t) \left[1 - \left(1 + \frac{ t}{\tau}\right) e^{- t/\tau}\right], 
\end{equation}

\begin{equation}
    \phi( x=\pm 1, y, t) = \pm \Xi_0 \frac{\sin( k_0 y)}{k_0}\left\{\left(\frac{t}{\tau}\right)^2e^{- t/\tau}\cos( \omega_0 t) - \omega_0\sin(\omega_0 t)\left[1-\left(1+\frac{ t}{\tau}\right) e^{- t/\tau}\right]\right\}, 
\end{equation}
with $\tau$ being a small growth parameter, following \cite{cole2004forced}. The code is periodic in the $y$ direction, where the Fast Fourier Transform is used to calculate spatial derivatives. The $x$ direction is non-periodic and a finite difference scheme of the sixth order is used to calculate spatial derivatives \cite{lele1992compact}. We use an explicit Runge-Kutta of the third order to advance in time. We impose the 2/3 rule for dealiasing with respect to the periodic variable, and a pseudo-spectral filter in $x$ \cite{lele1992compact}. Simulations are performed by using a grid $L_x\times L_y=[-1,1]\times[0,2\pi]$ with mesh points $N_x\times N_y=512\times32$ for $S=10^4$, $N_x\times N_y=1024\times32$ for $S=10^5$, $N_x\times N_y=2048\times16$ for $S=10^6$, and  $N_x\times N_y=4096\times16$ for $S=10^7$, with $N_x$  chosen to resolve their respective boundary layers. In all simulations we have fixed $\tau = 0.001$,  $\Xi_0 = 0.001$, and wavenumber $k_0 = 1$ while varying $S$ and the frequency $\omega_0$.

\bibliography{sample.bib}
\bibliographystyle{plain}
\bibliographystyle{unsrt}

\end{document}